\documentclass[12pt]{article}
\large

\title{ Renormalizable quantum
field theory as a limit of a quantum field model on the loop
space.}

\vspace{0.5cm}

\author{ Yu.P. Solovyov, V.V. Belokurov and E.T. Shavgulidze    \\
{\em Lomonosov Moscow State University, Russia }
\\ {\it e-mail: belokur@rector.msu.ru}}

\date{ \ \ \  }
\begin{document}
\maketitle

\begin{abstract}
A nonlocal generalization of quantum field theory in which
momentum space is the space of continuous maps of a circle into
$\mathbf{R}^4$ is proposed. Functional integrals in this theory
are proved to exist. Renormalized  quantum field model is obtained
as a local limit of the proposed theory.
\end{abstract}
\vspace{0.5cm}

 Consider a quantum field model defined in the following way. Let
the momentum space of the theory be the Banach space
$$
\mathcal{P}= C(S^1,\mathbf{R}^4)
$$
of all continuous maps of a circle of a unit length into
$\mathbf{R}^4$ with the norm
$$ \|p\| _\mathcal{P} =\max \limits _{ \tau \in S^1}\|p(\tau)\|\,,$$
where $\|\cdot \|$ is the canonical norm in $\mathbf{R}^4$.

An arbitrary element of this space can be represented in the form
\begin{equation}
   \label{1}
p(\tau)=r+\frac{1}{\sqrt{\lambda}}\xi(\tau)\,.
\end{equation}
Here $\xi(\tau)$ satisfies the condition
\begin{equation}
   \label{2}
\int \limits _{S^{1}}\, \,\xi(\tau)\, d\tau\,=\,0\,.
\end{equation}
We denote the space of maps $\xi$ as $C_{0}(S^1,\mathbf{R}^4)$.

Consider the group $G$ of three times differentiable
diffeomorphisms
$$
 G=Diff\,^{3}_{+}\left( S^{1}\right)\,,
$$
$$
g\in G \ \ \ \{g:\ S^{1}\longrightarrow S^{1}\,,\ \ \
g'(\tau)>0\}\,.
$$
Define the action of the group on the space $\mathcal{P}$ in the
following way:
\begin{equation}
   \label{3}
gp(\tau)=p\,\left(g^{-1}(\tau)\right)
\,\frac{1}{\sqrt{\left(g^{-1}\right)'(\tau)}}\,.
\end{equation}

Consider the Wiener measure on the space $\mathcal{P}$ with the
dispersion equal to $\sqrt{\lambda}$ and zero mathematical
expectation
\begin{equation}
   \label{4}
w_{\lambda}(dp)=\exp\left\{-\frac{\lambda}{2}\,\int \limits
_{S^{1}}\, \,\|p'(\tau)\|^{2}\, d\tau \right\}\,dp\,,\ \ \
\lambda>0\,.
\end{equation}
For any continuous bounded functional  $F(p)\,,$ that satisfies
the inequality $|F(p)|\leq \frac{C}{1+\|p\|^{5}} \,,$ the
following equation is valid
$$
\int \limits_{\mathcal{P}}\,F(p)\ w_{\lambda}(dp)= \int
\limits_{\mathbf{R}^4} \ \int
\limits_{C_{0}(S^1,\mathbf{R}^4)}\,F(r+\frac{1}{\sqrt{\lambda}}\xi)\
w_{1}(d\xi)\ dr\,.
$$
From here, the local limit of the integral follows
\begin{equation}
   \label{5}
\lim \limits_{\lambda\rightarrow +\infty}\int
\limits_{\mathcal{P}}\,F(p)\ w_{\lambda}(dp)= \int
\limits_{\mathbf{R}^4} \,F(r)\,dr\,.
\end{equation}

It is known that measures invariant with respect to the group $G$
do not exist. However, the measure $w_{\lambda}(dp)$ is
quasi-invariant. \cite{(Shavgulidze. 1978.)}

 It transforms as
\begin{equation}
   \label{6}
   w_{\lambda}(d\,(gp)\,)=\exp\left\{\frac{\lambda}{4}\,\int \limits
_{S^{1}}\,\mathcal{S}_{g}(\tau) \,\|p(\tau)\|^{2}\, d\tau
\right\}\ w_{\lambda}(dp)\,,
\end{equation}
Here $\mathcal{S}_{g}$ denotes the Schwarz derivative
\begin{equation}
   \label{7}
\mathcal{S}_{g}(\tau)= \frac{g'''(\tau)}{g'(\tau)}
-\frac{3}{2}\left(\frac{g''(\tau)}{g'(\tau)}\right)^2\,.
\end{equation}

Consider the space $E$ of all square-integrable over the Wiener
measure functions $\varphi :\mathcal{P}\to \mathbf{C}$, satisfying
the equation
$$
\varphi (p)=\overline{\varphi(-p)}
$$
for all $p\in \mathcal{P}$.

The space $E$ is a Hilbert space over the real field $\mathbf{R}$
with the scalar product
$$
(\varphi,\phi)_E= \int \limits_{\mathcal{P}}\,\varphi
(p)\overline{\phi (p)}\ w_{\lambda}(dp)\,.
$$

Functions $\varphi$ and $\phi$ realize a regular unitary
representation of the group $G$ in the Hilbert space $E$:
\begin{equation}
   \label{8}
   g\,\varphi (p)\,=\,\varphi (g\,p)\
\exp\left\{\frac{\lambda}{8}\,\int \limits
_{S^{1}}\,\mathcal{S}_{g}(\tau) \,\|p(\tau)\|^{2}\, d\tau
\right\}\,.
\end{equation}

Free action
\begin{equation}
   \label{9}
\mathcal{A}_0[\varphi ] =\int \limits_{\mathcal{P}} \,|\varphi (p)
|^2 \, \Omega ^{2}(p)\ w_{\lambda}(dp)\,
\end{equation}
with
\begin{equation}
   \label{10}
\Omega ^{2}(p)=\left(\int\limits_{S^1}\frac{d \tau}{\|p(\tau )
\|^2} \right)^{-1}+ m^2\,
\end{equation}
is invariant with respect to the group  $G$.

Note that
$$
\lim \limits_{\lambda\rightarrow +\infty}\,\mathcal{A}_0[\varphi
]= \int \limits_{\mathbf{R}^4} \,|\varphi (r) |^2\, \left(\|r \|^2
+ m^{2} \right)\, dr\,.
$$
Thus, in the limit case we get the action of free scalar field.

Let us construct the interaction term. In the limit case $(
\lambda\rightarrow +\infty) $ it should give the usual interaction
$\varphi^{4}$.

To get (quasi-) invariant expression we change  $p$ to $g\,p$ and
consider group averaging over $G$ with the quasi-invariant measure
\cite{(Shavgulidze. 2000.)}
\begin{equation}
   \label{11}
\mu_{\beta}(dg)=\exp\left\{-\frac{\beta}{2}\,\int \limits
_{S^{1}}\,
\,\left[\left(\frac{g''(\tau)}{g'(\tau)}\right)'\right]^{2}\,
d\tau \right\}\,dg\,,
\end{equation}
where $\beta$  is an arbitrary positive parameter.

It is convenient to write the parameter $\beta $ in the form
$\beta=\alpha\lambda \,$ with an arbitrary positive $\alpha$.

Taking into account the transformation rules for the field and the
measure we propose the following formula for interaction
$$
\mathcal{A}_1[\varphi ] =\int\limits_{\mathcal{P}}\cdots
\int\limits_{\mathcal{P}}\, \varphi (p_1)\varphi (p_2)\varphi
(p_3)\varphi (p_4)
$$
$$
\int\limits_{S^1}\delta ( p_1(\tau_1) - p_5(\tau_1))\, \|
p_1(\tau_1)\|^2 d \tau_1 \, \int\limits_{S^1}\frac{ d
\tau}{\|p_1(\tau ) \|^2}\,
$$
$$
 \int\limits_{S^1}\delta ( p_2(\tau_2)
- p_6(\tau_2))\, \| p_2(\tau_2)\|^2 d \tau_2\,
\int\limits_{S^1}\frac{d \tau}{\|p_2(\tau ) \|^2}\,
$$
$$
\int\limits_{S^1}\delta ( p_3(\tau_3) - p_7(\tau_3))\, \|
p_3(\tau_3)\|^2 d \tau_3 \, \int\limits_{S^1}\frac{d
\tau}{\|p_3(\tau ) \|^2}\,
$$
$$
 \int\limits_{S^1}\delta ( p_4(\tau_4) - p_8(\tau_4)) \, \|
p_4(\tau_4)\|^2 d \tau_4\, \int\limits_{S^1}\frac{d
\tau}{\|p_4(\tau ) \|^2}\,
$$
$$
\int\limits_{S^1}\delta ( p_5(\tau_5)+p_6(\tau_5)+ p_7(\tau_5) +
p_8(\tau_5)) \, \|p_5(\tau_5) \|^2 d \tau_5
\int\limits_{S^1}\frac{d \tau}{\|p_5(\tau) \|^2}
$$
$$
\int\limits_{G}\,\exp\left\{   \frac{\lambda}{8}
\int\limits_{S^1}\,\mathcal{S}_{g_{1}}(\tau)\,\|p_1(\tau ) \|^2\,
d\tau \right\} \ \mu_{\alpha \lambda}( dg_1) \ w_{\lambda}(dp_1)
$$
$$
\cdots
$$
$$
\int\limits_{G}\,\exp\left\{   \frac{\lambda}{8}
\int\limits_{S^1}\,\mathcal{S}_{g_{8}}(\tau)\,\|p_8(\tau ) \|^2\,
d\tau \right\} \ \mu_{\alpha \lambda}( dg_8) \ w_{\lambda}(dp_8)
$$
We see that interaction in this model is nonlocal.

Using the following notations
$$ \Delta (p_1, p_5)\equiv
\int\limits_{\mathbf{S^1}}\frac{ d \tau}{\|p_1(\tau ) \|^2}\,
\int\limits_{\mathbf{S^1}}\delta \left( p_1(\tau_1) -
p_5(\tau_1)\right) \| p_1(\tau_1)\|^2 d \tau_1 \,,
$$
$$
\mathcal{U}_{\alpha\lambda}(p)\equiv \int\limits_{G}\,\exp\left\{
\frac{\lambda}{8}
\int\limits_{\mathbf{S^1}}\,\mathcal{S}_{g}(\tau)\,\|p(\tau )
\|^2\, d\tau \right\} \ \mu_{\alpha \lambda}( dg) \,,
$$
we write it in a more compact form
$$
\mathcal{A}_1[\varphi ] =\int\limits_{\mathcal{P}}\cdots
\int\limits_{\mathcal{P}}\, \varphi
(p_1)\,\mathcal{U}_{\alpha\lambda}(p_1)\,\Delta (p_1, p_5)\
$$
$$
\varphi (p_2)\,\mathcal{U}_{\alpha\lambda}(p_2)\,\Delta (p_2,
p_6)\
$$
$$ \varphi
(p_3)\,\mathcal{U}_{\alpha\lambda}(p_3)\,\Delta (p_3, p_7)\
$$
$$
 \varphi
(p_4)\,\mathcal{U}_{\alpha\lambda}(p_4)\,\Delta (p_4, p_8)\
$$
$$
\Delta (p_5, -p_6-p_7-p_8)
$$
$$
\mathcal{U}_{\alpha\lambda}(p_5)\,\mathcal{U}_{\alpha\lambda}(p_6)\,\mathcal{U}_{\alpha\lambda}(p_7)\,
\mathcal{U}_{\alpha\lambda}(p_8)\ $$
$$ w_{\lambda}(dp_1)\,\cdots
\, w_{\lambda}(dp_8)\,.
$$

Getting in mind the effects of renormalization we consider also
the following addition to the action
$$
\mathcal{A}_2[\varphi ] = \int\limits_{\mathcal{P}}
\int\limits_{\mathcal{P}}\int\limits_{\mathcal{P}}\int\limits_{\mathcal{P}}\,
\varphi (p_1)\,\varphi (p_2)
$$
$$
 \Delta\left(p_1,p_3\right)\
\Delta\left(p_2,p_4\right)\ \Delta\left(p_3,-p_4\right)
$$
$$
\mathcal{U}_{\alpha \lambda}( p_1) \ w_{\lambda}(dp_1) \
\mathcal{U}_{\alpha \lambda}( p_2) \ w_{\lambda}(dp_2)
$$
$$ \mathcal{U}_{\alpha \lambda}( p_3) \ w_{\lambda}(dp_3)\
\mathcal{U}_{\alpha \lambda}( p_4) \ w_{\lambda}(dp_4)\,.$$

Now the total action is given by the equation
$$
\mathcal{A}[\psi ]=\mathcal{A}_0[\psi ]+ \kappa_1
\mathcal{A}_1[\psi ]+\kappa_2 \mathcal{A}_2[\psi ]\,.
$$

The following theorem is valid

 \textbf{Theorem 1.}

   {\it  For any positive  $\kappa _1$ and any
   $\psi \in E$ there exists the integral}
\begin{equation}
   \label{12}
      \int\limits_{E} e^{-\mathcal{A}[\varphi]-i(\varphi,\psi)_E}d\varphi\,.
\end{equation}

To prove this theorem note that for any $\varphi \in E$ the
functional $ \mathcal{A}_1 [\varphi ]$ is nonnegative. Therefore,
the functional $ e^{-\kappa_1 \mathcal{A}_1[\varphi ]-\kappa_2
\mathcal{A}_2[\varphi ] -i(\varphi,\psi)_E}$ is bounded on $E$.
Hence, to prove that the integral (\ref{12}) converges it is
sufficient to verify the existence of the integrals
   $$\int_E \mathcal{A}_1[\psi]e^{-\mathcal{A}_0[\psi]}d\psi$$
   and
   $$\int_E \mathcal{A}_2[\psi]e^{-\mathcal{A}_0[\psi]}d\psi\ $$
   \cite{(Smolyanov and Shavgulidze. Continual integrals. 1990.)}.

It can be done with the help of several bounds including those
that are given by the lemmas.

\textbf{Lemma 1.} {\it There exists $c_1>0$ such that $\forall \ r
\in \mathbf{R}^4$ and $\forall \ t_1,t_2 \in \mathbf{S^1}\
(t_1\neq t_2)$ the following inequality is valid}
$$
 \int\limits_{\mathcal{P}}\, \delta \left( p(t_1) - r\right) \|
p(t_1)\|^2 \, \| p(t_2)\|^2  \int\limits_{\mathbf{S^1}}\frac{d
\tau}{\|p(\tau ) \|^2}\ \mathcal{U}_{\alpha\lambda}(p) \
w_{\lambda}(dp) \leq c_1\left(1+r^{4}\right)\,.
$$

\textbf{Lemma 2.} {\it There exists $\ c_2>0$ such that $\forall \
t_1,t_2 \in \mathbf{S^1}\ (t_1\neq t_2)$ the following inequality
is fulfilled}
$$
 \int\limits_{\mathcal{P}}\, \left[\Omega
^{2}(p_{1})\right]^{-1}\, \| p(t_1)\|^2 \, \left( \|
p(t_2)\|^2+1\right)\,  \left(\int\limits_{\mathbf{S^1}}\frac{d
\tau}{\|p(\tau ) \|^2}\right)^{2}\,
\left(\mathcal{U}_{\alpha\lambda}(p)\right)^{2} \ w_{\lambda}(dp)
\leq c_2\,.
$$

We have already discussed that when $\lambda\rightarrow +\infty$
the set of measures $ w_{\lambda}$ converges in a weak sence to
Lebesgue measure on $ \mathbf{R}^4 \subset\mathcal{P}\,.$

The limit of $ \mathcal{A}_0[\phi_{\varphi} ]$ gives the action of
free scalar field.

Similarly, when $\alpha\rightarrow +\infty$ и $\lambda\rightarrow
+\infty\,$, the measure $\ \mathcal{U}_{\alpha \lambda}( p) \
w_{\lambda}(dp)$ converges in a weak sence to a Lebesgue measure.
And for the limits of $\mathcal{A}_1[\phi_{\varphi} ]$ and
$\mathcal{A}_2[\phi_{\varphi} ]$ we have the action of
$\varphi^{4}$ model.
$$
\lim \limits_{\lambda\rightarrow +\infty\,,\alpha\rightarrow
+\infty}\,\mathcal{A}_1[\phi_{\varphi} ]=\int
\limits_{\mathbf{R}^4} \int \limits_{\mathbf{R}^4}\int
\limits_{\mathbf{R}^4}\int
\limits_{\mathbf{R}^4}\,\delta\left(r_{1}+r_{2}+r_{3}+r_{4}\right)\,\varphi(r_{1})
\varphi(r_{3})\varphi(r_{3})\varphi(r_{4})\,dr_{1}dr_{2}dr_{3}dr_{4}\,.
$$
$$
\lim \limits_{\lambda\rightarrow +\infty\,,\alpha\rightarrow
+\infty}\,\mathcal{A}_2[\varphi ]=\int \limits_{\mathbf{R}^4}
\,|\varphi (r) |^2\,  dr\,.
$$

It can be proven that the local limit of our model is the
renormalized $\varphi^{4}$ theory. That is, the following theorem
is valid.

\textbf{Theorem 2.}

   {\it There are functions $\kappa _1(\theta,\alpha)$ and
$\kappa _2(\theta,\alpha)$ such that for all integer $n\geq 0$ the
following limits exist}
$$
\lim \limits _{ \lambda \to +\infty} \lim \limits _{ \alpha \to
+\infty} \frac{\partial^n }{\partial \theta^n}\left.
\frac{\int\limits_{E} e^{-\mathcal{A}_0[\psi]- \kappa
_1(\theta,\alpha)\mathcal{A}_1[\psi]- \kappa
_2(\theta,\alpha)\mathcal{A}_2[\psi]
-i(\psi,\phi_{\varphi})_E}\,d\psi} {\int\limits_{E}
e^{-\mathcal{A}_0[\psi]- \kappa
_1(\theta,\alpha)\mathcal{A}_1[\psi]- \kappa
_2(\theta,\alpha)\mathcal{A}_2[\psi]}\,d\psi}\right|_{\theta=0}\,.
$$

To prove this theorem we make a substitution
\begin{equation}
   \label{2.1}
f(\tau)=\sqrt{\lambda}\left(\frac{g''(\tau)}{g'(\tau)}\right)\,.
\end{equation}

The measure $\mu$ looks like
\begin{equation}
   \label{2.2}
\mu_{\alpha \lambda}(dg)=\exp\left\{-\frac{\alpha}{2}\,\int
\limits _{S^{1}}\, \,\left(f'(\tau)\right)^{2}\, d\tau
\right\}\,df= w_{\alpha}(df)\,.
\end{equation}

Schwarz derivative $\mathcal{S}_{g}$ takes the form
\begin{equation}
   \label{2.3}
\mathcal{S}_{g}(\tau)=\frac{1}{\sqrt{\lambda}}\,f'(\tau)-\frac{1}{2\lambda}\,f^{2}(\tau)\,.
\end{equation}

Also, we have
\begin{equation}
   \label{2.4}
\lim \limits_{\lambda\rightarrow +\infty}\exp\left\{
\frac{\lambda}{8}
\int\limits_{S^1}\,\mathcal{S}_{g}(\tau)\,\|p(\tau ) \|^2\, d\tau
\right\}=
\end{equation}
$$
\exp\left\{ -\frac{1}{16}
\int\limits_{S^1}\,f^{2}(\tau)\,d\tau\,\|r\|^{2} + \frac{1}{4}
\int\limits_{S^1}\,f'(\tau)\,\left(r,\xi(\tau ) \right)\,
d\tau\right\}\,.
$$

Now, for the limit $\lim \limits_{\lambda\rightarrow +\infty\,,\
\alpha=const} \mathcal{A}_1$ we get
$$
\lim \limits_{\lambda\rightarrow +\infty\,,\ \alpha=const}
\mathcal{A}_1[\varphi ] =\int\limits_{\mathbf{R}^4} ...
\int\limits_{\mathbf{R}^4}\,\int\limits_{F} ...
\int\limits_{F}\,\int\limits_{C_{0}(S^1,\mathbf{R}^4)} ...
\int\limits_{C_{0}(S^1,\mathbf{R}^4)}\, \varphi (r_1)\varphi
(r_2)\varphi (r_3)\varphi (r_4)
$$
$$
\delta ( r_1 - r_5)  \,
 \delta ( r_2
- r_6) \, \delta ( r_3 - r_7)  \, \delta ( r_4 - r_8) \, \delta (
r_5+r_6+ r_7 + r_8)
$$
$$
\int\limits_{}\,\exp\left\{ -\frac{1}{16}
\int\limits_{S^1}\,f^{2}_{1}(\tau)\,d\tau\,\|r\|^{2}_{1} +
\frac{1}{4}
\int\limits_{S^1}\,f'_{1}(\tau)\,\left(r_{1},\xi_{1}(\tau )
\right)\, d\tau\right\} \
w_{\alpha}(df_1)\,w_{1}(d\xi_{1})\,dr_{1}
$$
$$
\cdot\,\cdot\,\cdot
$$
$$
\int\limits_{}\,\exp\left\{ -\frac{1}{16}
\int\limits_{S^1}\,f^{2}_{8}(\tau)\,d\tau\,\|r\|^{2}_{8} +
\frac{1}{4}
\int\limits_{S^1}\,f'_{8}(\tau)\,\left(r_{8},\xi_{8}(\tau )
\right)\, d\tau\right\} \
w_{\alpha}(df_8)\,w_{1}(d\xi_{8})\,dr_{8}\,.
$$

Integrations over $r_{5},...,r_{8}$ and $\xi_{1},...,\xi_{8}$
result in
$$
\lim \limits_{\lambda\rightarrow +\infty\,,\ \alpha=const}
\mathcal{A}_1[\varphi ] =\int\limits_{\mathbf{R}^4} ...
\int\limits_{\mathbf{R}^4} \varphi (r_1)\varphi (r_2)\varphi
(r_3)\varphi (r_4) \
 \int\limits_{F}...\int\limits_{F}
$$
$$
\exp\left\{ -\frac{1}{32}
\int\limits_{S^1}\,\left[f^{2}_{1}(\tau)+f^{2}_{5}(\tau)\right]\,d\tau\
\|r_1\|^{2} \right\} \, \exp\left\{ -\frac{1}{32}
\int\limits_{S^1}\,\left[f^{2}_{2}(\tau)+f^{2}_{6}(\tau)\right]\,d\tau\
\|r_2\|^{2} \right\}
$$
$$
\exp\left\{ -\frac{1}{32}
\int\limits_{S^1}\,\left[f^{2}_{3}(\tau)+f^{2}_{7}(\tau)\right]\,d\tau\
\|r_3\|^{2} \right\}  \exp\left\{ -\frac{1}{32}
\int\limits_{S^1}\,\left[f^{2}_{4}(\tau)+f^{2}_{8}(\tau)\right]\,d\tau\
\|r_4\|^{2} \right\}
$$
$$
w_{\alpha}(df_1)\,\cdot\cdot\cdot\, w_{\alpha}(df_8)\ \ \delta (
r_1+r_2+ r_3 + r_4)\,dr_{1}\,\cdot\cdot\cdot\,dr_{4}\,.
$$

Finally, evaluating functional integrals over $f_i$ we obtain
\begin{equation}
   \label{2.5}
\lim \limits_{\lambda\rightarrow +\infty\,,\ \alpha=const}
\mathcal{A}_1[\varphi ] =
\end{equation}
$$
\int\limits_{\mathbf{R}^4} ... \int\limits_{\mathbf{R}^4} \varphi
(r_1)\,I\left(\alpha\,,\,\frac{1}{4}\|r_1\|\right)\cdots\varphi
(r_4)\,I\left(\alpha\,,\,\frac{1}{4}\|r_4\|\right) \
$$
$$
 \delta ( r_1+r_2+ r_3 + r_4)\,dr_{1}\,\cdot\cdot\cdot\,dr_{4}\,.
$$
Here
\begin{equation}
   \label{2.6}
I(\alpha,a)=\frac{a}{\sqrt{\alpha}}\,\frac{\exp\left(-\frac{a}{2\sqrt{\alpha}}
\right)}{1-\exp\left(-\frac{a}{\sqrt{\alpha}}\right)}\,.
\end{equation}

Similarly, for $\ \ \lim \limits_{\lambda\rightarrow +\infty\,,\
\alpha=const} \mathcal{A}_2\ \ $ we obtain
\begin{equation}
   \label{2.7}
\lim \limits_{\lambda\rightarrow +\infty\,,\ \alpha=const}
\mathcal{A}_2[\varphi ] =
\end{equation}
$$
\int\limits_{\mathbf{R}^4}\, \int\limits_{\mathbf{R}^4} \varphi
(r_1)\,I\left(\alpha\,,\,\frac{1}{4}\|r_1\|\right)\ \varphi
(r_2)\,I\left(\alpha\,,\,\frac{1}{4}\|r_2\|\right) \
$$
$$
 \delta ( r_1+r_2)\,dr_{1}\,dr_{2}\,.
$$
Note that the factors $I$ decrease  very fast at large $\|r\|$.
Therefore, we can consider the above formulae  as a regularization
of $\varphi^{4}\,$ model with the regularization parameter
$\alpha$. It belongs to a set of regularizations that can be used
for the proof of renormalizability of $\varphi^{4}\,$ model
\cite{(Zavyalov. 1979.)}.

In a similar manner, it is possible to prove the existence of the
functional integral in $\varphi^{4}$ model in 3-dimensional
space-time
$$
\lim \limits _{ \lambda \to +\infty} \lim \limits _{ \alpha \to
+\infty}$$ $$ \frac{\int\limits_{E} e^{-\mathcal{A}_0[\varphi]-
\kappa_1\mathcal{A}_1[\varphi]-
\kappa_2(\theta,\alpha)\mathcal{A}_2[\varphi]
-i(\varphi,\chi)_E}\,d\varphi} {\int\limits_{E}
e^{-\mathcal{A}_0[\varphi]- \kappa _1\mathcal{A}_1[\varphi]-
\kappa _2(\theta,\alpha)\mathcal{A}_2[\varphi]}\,d\varphi}\,.
$$

\end{document}